\documentclass[twocolumn,showpacs,preprintnumbers,prb]{revtex4}

\usepackage{graphicx}
\usepackage{dcolumn}
\usepackage{bm}
\usepackage{bbm}
\usepackage{psfrag}
\usepackage{multirow,amssymb,amsbsy,amsmath}
\usepackage{stmaryrd}

\newcommand{\I}{i}
\newcommand{\E}{e}
\newcommand{\D}{d}

\newcommand{\vek}[1]{\bm{#1}}
\newcommand{\dod}[2]{\frac{\D #1}{\D #2}}

\newcommand{\pop}[2]{\frac{\partial #1}{\partial #2}}

\newcommand{\Funktion}[2]{#1\kern-0.2em\left(#2\right)}

\newcommand{\trtxt}[2][]{\text{Tr}_{#1}\{#2\}}

\newcommand*{\ket}[1]{\mathopen{|}#1\mathclose{\rangle}}
\newcommand*{\braket}[2]{\mathopen{\langle}#1|#2\mathclose{\rangle}}

\newcommand*{\expvalshort}[1]{\mathopen{\langle} #1 \mathclose{\rangle}}
\newcommand*{\expval}[3]{\mathopen{\langle}#1|#2|#3\mathclose{\rangle}}
\newcommand{\refsec}[1]{Sect.~\ref{#1}}
\newcommand{\refeq}[1]{Eq.~(\ref{#1})}
\newcommand{\reffig}[1]{Fig.~\ref{#1}}

\newcommand{\comutxt}[3][]{[#2,#3]_{#1}}

\newcommand{\Hamil}[1][]{\hat{H}_{\text{#1}}}
\newcommand{\Lop}[1][L]{\mathcal{#1}}
\newcommand{\dens}{\hat{\rho}}
\newcommand{\Pauli}[1]{\hat{\sigma}_{#1}}
\newcommand{\one}{\hat{1}}

\begin{document}

\preprint{APS/123-QED}

\title{Transport in open spin chains: a Monte Carlo wave-function approach}

\author{Mathias Michel}
\email{m.michel@surrey.ac.uk}
\affiliation{
	Advanced Technology Institute, %
	Faculty of Engineering and Physical Sciences, %
	University of Surrey, %
	Guildford, %
	GU2 7XH, %
	United Kingdom}
\author{Ortwin Hess}
\affiliation{
	Advanced Technology Institute, %
	Faculty of Engineering and Physical Sciences, %
	University of Surrey, %
	Guildford, %
	GU2 7XH, %
	United Kingdom}
\author{Hannu Wichterich}
\email{hwichter@uni-osnabrueck.de}
\affiliation{
	Physics Department, %
	University of Osnabr\"uck, %
	Barbarastr.\ 7, %
	49069 Osnabr\"uck, %
	Germany}
\author{Jochen Gemmer}
\affiliation{
	Physics Department, %
	University of Osnabr\"uck, %
	Barbarastr.\ 7, %
	49069 Osnabr\"uck, %
	Germany}%

\date{\today}

\begin{abstract}
We investigate energy transport in several two-level atom or spin-1/2 models by a direct coupling to heat baths of different temperatures.
The analysis is carried out on the basis of a recently derived quantum master equation which describes the nonequilibrium properties of internally weakly coupled systems appropriately.
For the computation of the stationary state of the dynamical equations, we employ a Monte Carlo wave-function approach.
The analysis directly indicates normal diffusive or ballistic transport in finite models and hints toward an extrapolation of the transport behavior of infinite models.
\end{abstract}

\pacs{05.60.Gg, 44.10.+i, 05.70.Ln}

\maketitle

%
%

%
%
\section{Introduction}
\label{sec:1}

The transport of energy or heat has been intensively studied since Fourier introduced his famous law of heat conduction in 1807.
Surprisingly, still 200 years later, some fundamental problems remain unsolved\cite{Bonetto2000}.
Contrary to our everyday experience, the appearance of diffusive behavior according to Fourier's famous law is difficult to obtain from the direction of any underlying microscopic theory.

A question of central relevance concerns the classification of the transport properties of a system into \emph{normal diffusive} or \emph{ballistic} behavior.
In the classical domain, it seems to be largely accepted that normal energy transport, i.e., spatial diffusion instead of ballistic transport or localization, requires the chaotic dynamics of a nonintegrable system \cite{Lepri2003,Vollmer2002}.
In the quantum regime, however, the question whether diffusive behavior follows from the underlying theory turns out to be a controversial issue \cite{Zotos1997,Jung2006,Michel2005,Michel2005II,Michel2006II}.
This is mostly due to the nontrivial character of the question, how energy is transported on the microscopic scale.

Amongst the many different techniques of investigating transport in quantum mechanics, let us consider two approaches in more detail here.
The first one is the prominent Green-Kubo formula.
Derived on the basis of linear response theory it has originally been formulated for electrical transport\cite{Kubo1957,Kubo1991,Mori1956}.
Therein, the system is perturbed by an external force, first the electric field, applied to the system.
The resulting current of charge is viewed as the response to this external perturbation.
Finally, the transport coefficient (conductivity) follows from a current-current autocorrelation.
The same approach is also used in the different case of density driven transport, e.g., the transport of energy or heat.
In such a situation, the current is driven by a much more complicated mechanism (the coupling of reservoirs to the system) than simply an external force, which can be nicely written as a term within the Hamiltonian of the system.
Nevertheless, the correlation function is ad hoc transferred to the density driven scenario by replacing the electrical current by the energy or heat current\cite{Luttinger1964}.
However, the justification of this replacement remains a conceptual problem here.

One big advantage of this widely used approach is certainly its computability after having diagonalized the system's Hamiltonian.
A nice overview of results from the Kubo formula for spin models can be found, e.g., in the work by Heidrich-Meisner\cite{Heidrich2005} (for further reading we suggest the comprehensive literature cited therein).
However, in most cases, a direct analytical solution for an infinite system is not feasible and the interpretation of the results for finite systems seems to be not straight forward.
For a finite system, the frequency dependent transport coefficient consists of numerous delta peaks with different weights at frequencies $\omega$ and is zero elsewhere.
How to extract the dc-conductivity (interesting especially for the energy transport) of the finite system from this result or extrapolate the conductivity for the infinite one is a difficult question\cite{Zotos2004,Heidrich2003,Heidrich2005II,Gemmer2006II}.

A different approach to investigate the transport behavior of a system is more connected to the experimental measurement of heat conductivities:
The system is directly coupled to heat baths of different temperatures within the theory of open quantum systems\cite{Saito2003,Saito2000,Michel2003,Gemmer2004,Michel2006III}.
That means that the Liouville von Neumann equation, describing the time evolution of the density operator of the system, is extended by incoherent damping terms simulating the influence of the heat baths.
How to set up the correct dynamical equation here is highly nontrivial and involves the combination of many subtle approximation schemes.
In the case of an improper approach, the derivation can lead to mathematically correct, but physically irrelevant dynamical equations as discussed recently\cite{Wichterich2007}.
Having derived a proper quantum master equation (QME), the interpretation of the results for finite systems is relatively easy:
After finding the stationary state of the dissipative dynamics, all interesting quantities as currents and energy profiles are simply accessible by computing the expectation value of the respective operator.
However, also this approach is restricted to finite systems since a complete analytical solution for larger systems is not available.
Thus, the extrapolation to infinite systems needs a careful discussion to exclude errors due to the finite size of the investigated models as well.

In the present paper, we will consider several model systems according to their transport properties.
This is mainly done by the bath coupling method as discussed above, and by comparing with results from the Kubo formula.
Let us start in \refsec{sec:2} with an introduction to the QME, the necessary observables and the Monte Carlo wave-function technique\cite{Breuer2002,Plenio1998} which is used to integrate the QME.
Afterwards, we will present the results for several model systems: for chainlike systems in \refsec{sec:3} and more complex ones in \refsec{sec:4}, followed by our summary and conclusion.

%
%
\section{Background}
\label{sec:2}

%
\subsection{Model system}
\label{sec:2:1}

The considered system consists of $N$ weakly coupled subunits described by the Hamiltonian
\begin{equation}
	\label{eq:1}
	\Hamil 
	= \Hamil[loc] + \Hamil[int]
	= \sum_{\mu=1}^N \hat{h}^{(\mu)} + J\sum_{\mu=1}^{N-1} \hat{h}^{(\mu,\mu+1)}\,.
\end{equation}
The first part $\Hamil[loc]$ of the Hamiltonian contains the local spectra of the subunits.
The second part $\Hamil[int]$ describes the interaction between adjacent sites with the coupling strength $J$.
Here, we require that the interaction is weak in the sense that the energy contained in the local part is much larger than the energy contained in the interaction part, $\langle\Hamil[loc]\rangle \gg \langle\Hamil[int]\rangle$.

More concretely, we will investigate one-dimensional (1D) or quasi-1D chains of two-level atoms or spin-1/2 particles.
Both, two-level atoms and spins, are described by the same algebra, and thus, it is convenient to use the Pauli operators $\{\one,\Pauli{x},\Pauli{y},\Pauli{z}\}$ as a suitable operator basis here.
The above mentioned weak coupling claim is fullfilled by introducing a local Zeeman splitting $\frac{\Omega}{2}\,\Pauli{z}$, where we require that $\Omega$ is much larger than the coupling constant $J$.
Note that this \emph{weak internal coupling} constraint is a necessary precondition for the validity of the master equation introduced below, i.e., we are not able to consider systems with $\Omega$ approaching the same magnitude as $J$ here.
Hence, the models described by the Hamiltonian given in \refeq{eq:1} are not to be confused with spin chains in cuprates, e.g., where the local field is always close to zero, and thus, small compared to the coupling strength.
However, the discussed models can be seen as spin chains in strong external fields, or simply as weakly coupled chains of two-level atoms as frequently considered in quantum optics and quantum information theory.

To investigate the transport properties of these systems, they will be explicitly coupled to independent environments of different temperatures.
Let us discuss the appropriate  (QME)\cite{Breuer2002} to describe this situation in the following Section.

%
\subsection{Lindblad Quantum Master Equation}
\label{sec:2:2}

In general, the derivation of the QME from a microscopic model\cite{Breuer2002} (a system coupled to an infinitely large environment) relies on some well known approximation schemes the Markov\cite{Davies1974II,Davies1976} assumption, the Born approximation and the secular approximation\cite{Davies1974II,Duemcke1979}.
Recently, there was a discussion on how to derive a suitable Lindblad\cite{Gorini1976,Lindblad1976} QME in a nonequilibrium scenario\cite{Wichterich2007}, i.e., an equation to investigate transport in weakly coupled quantum systems.
The Lindblad form of a QME defines a trace and hermiticity preserving, completely positive dynamical map\cite{Breuer2002,Alicki1986}, which thus retains all properties of the density operator at all times.
In order to approach this dynamical equation the approximations are carefully carried out in a minimally invasive manner, to retain the central nonequilibrium properties of the model.
It was shown that this \emph{nonequilibrium} Lindblad QME is in very good accord with the results of the Redfield master\cite{Breuer2002} equation (non-Lindbladian), contrary to the standard Lindblad QME in the weak coupling limit\cite{Wichterich2007}.

In a nonequilibrium investigation, one needs two heat baths at different temperatures \emph{locally} coupled to the system, i.e., the heat baths couple only to a subunit at the edge of the system.
The QME of such a situation yields
\begin{equation}
	\label{eq:2}
	\dod{\dens}{t} = -\I\comutxt{\Hamil}{\dens} + \Lop[D]_{L}(\dens) + \Lop[D]_{R}(\dens)\,,
\end{equation}
where the dissipator $\Lop[D]_{L}$ refers to the left heat bath and $\Lop[D]_{R}$ to the right one, depending on the full density operator $\hat{\rho}$ of the system, i.e., the state of the chain described by the Hamiltonian (\ref{eq:1}).
Both dissipators depend on the coupling strength $\lambda$ between system and bath as well as the temperature of the bath, respectively.
Besides those incoherent damping terms, Eq.~(\ref{eq:2}) contains a coherent part containing the Hamiltonian  (\ref{eq:1}) of the system.

The dissipator describing the heat bath coupled to a subunit at the edge of the system yields
\begin{equation}
	\label{eq:3}
	\Lop[D]_{F}(\dens)
	= \sum_{k,l=1}^{2} (\gamma_F)_{kl}
	  \Big(\hat{F}_k\dens\hat{F}_l^{\dagger}
	- \frac{1}{2}\comutxt[+]{\hat{F}_l^{\dagger}\hat{F}_k}{\dens}\Big)
\end{equation}
with $F=L$ for the left and $F=R$ for the right heat bath.
The Lindblad operators $\hat{F}_k$ are given by
\begin{subequations}
\begin{align}
	\label{eq:4a}
	\hat{L}_1 &= \Pauli{+}^{(1)}\otimes\one^{(2)}\otimes\cdots\otimes\one^{(N)}\,,\\
	\label{eq:4b}
	\hat{L}_2 &= \Pauli{-}^{(1)}\otimes\one^{(2)}\otimes\cdots\otimes\one^{(N)}\,,\\
	\label{eq:4c}
	\hat{R}_1 &= \one^{(1)}\otimes\cdots\otimes\one^{(N-1)}\otimes\Pauli{+}^{(N)}\,,\\
	\label{eq:4d}
	\hat{R}_2 &= \one^{(1)}\otimes\cdots\otimes\one^{(N-1)}\otimes\Pauli{-}^{(N)}\,,
\end{align}
\end{subequations}
with the creation and annihilation operators $\Pauli{\pm}$.
Here, the operators given in Eqs.~(\ref{eq:4a}) and (\ref{eq:4b}) belong to the left bath and those in Eqs.~(\ref{eq:4c}) and (\ref{eq:4d}) to the right one.
The coefficient matrices depend on the respective bath temperature $\beta_F$ and are defined as
\begin{equation}
	\label{eq:5}
	\gamma_F = 
	\begin{pmatrix}
		\Gamma_F(\Omega) & \sqrt{\Gamma_F(\Omega)\Gamma_F(-\Omega)}\\
		\sqrt{\Gamma_F(\Omega)\Gamma_F(-\Omega)} & \Gamma_F(-\Omega)
	\end{pmatrix}\,
\end{equation}
according to the rates
\begin{equation}
	\label{eq:6}
	\Gamma_F(\Omega) = \frac{\lambda\Omega}{\E^{\beta_F\Omega}-1}\,
\end{equation}
with the bath coupling strength $\lambda$.
This concrete form of the $\gamma_F$-matrices follows from a phenomenological ansatz for the spectral density of the environment, here chosen to be of Ohmic\cite{Breuer2002,May2003} kind.

A remarkable property of Eq.~(\ref{eq:3}) is that it can be brought into Lindblad form by diagonalizing the coefficient matrices $\gamma_F$.
The complete dissipative part of Eq.~(\ref{eq:2}) then reads
\begin{equation}
	\label{eq:10}
	\Lop[D](\dens)
	= \sum_{k=1}^{4} \alpha_{k}
	\Big(\hat{E}_k\dens\hat{E}_k^{\dagger} -\frac{1}{2} \comutxt[+]{\hat{E}_k^{\dagger}\hat{E}_k}{\dens}\Big)
\end{equation}
with $\hat{E}_k$ being linear combinations of the operators $\hat{F}_k$ defined in Eqs.~(\ref{eq:4a})-(\ref{eq:4d}) and $\alpha_{k}$ being non-negative numbers.

That it is indeed possible to derive a Lindbladian QME is very important here, since a standard stochastic unravelling of this special type of equation is feasible.
Although extended stochastic schemes exist for the solution of general QMEs such as, e.g., the Redfield equation \cite{Breuer1999,Breuer2004,Breuer2004II}, these methods have turned out to be less efficient, in general, than the standard approach.

%
\subsection{Observables and Fourier's Law}
\label{sec:2:3}

The most interesting state of a nonequilibrium scenario is the local equilibrium state, i.e., the stationary state of the QME (\ref{eq:2}).
This state can be characterized by two central observables---the energy gradient and the energy current.
Let us use
\begin{equation}
	\label{eq:11}
	h^{(\mu)} = \trtxt{\hat{h}^{(\mu)}\hat{\rho}(t)}\,
\end{equation}
as a local energy density at site $\mu$ with $\hat{\rho}(t)$ being the state of the system at time $t$.
Since we are investigating internally weakly coupled subunits in the limit $\Omega\gg J$ the local energy density is approximated by the local Hamiltonian here.
Therefore, we neglect completely the contributions to the local energy by the interaction.
However, due to the smallness of $J$, these parts would be very small contributions to the above given energy density, and thus, would not dramatically change the results.

In order to obtain a current operator between two adjacent sites in the system, we consider the time evolution of the local energy operator given by the Heisenberg equation of motion for operators at site $\mu$
\begin{equation}
	\label{eq:12}
	\dod{}{t}\hat{h}^{(\mu)} = \I\comutxt{\hat{H}}{\hat{h}^{(\mu)}} + \pop{}{t}\hat{h}^{(\mu)}\,.
\end{equation}
Since $\hat{h}^{(\mu)}$ is not explicitly time dependent the last term vanishes.
Inserting Eq.~(\ref{eq:1}) into (\ref{eq:12}) yields 
\begin{equation}
	\label{eq:13}
	\dod{}{t}\hat{h}^{(\mu)} 
	= \I\Big(\comutxt{\hat{h}^{(\mu-1,\mu)}}{\hat{h}^{(\mu)}} +
		 \comutxt{\hat{h}^{(\mu,\mu+1)}}{\hat{h}^{(\mu)}}\Big)\,.
\end{equation}
Assuming that the local energy is a conserved quantity which is justified when $\Omega\gg J$ a discretized version of the continuity equation yields
\begin{equation}
	\label{eq:14}
	\dod{}{t}\hat{h}^{(\mu)}
	= \text{div} \hat{J}
	= \hat{J}^{(\mu,\mu+1)}-\hat{J}^{(\mu-1,\mu)}\,.
\end{equation}
By comparing Eq.~(\ref{eq:13}) and Eq.~(\ref{eq:14}) we find for the current operator
\begin{equation}
	\label{eq:15}
	\hat{J}^{(\mu,\mu+1)} = \I\comutxt{\hat{h}^{(\mu,\mu+1)}}{\hat{h}^{(\mu)}}\,.
\end{equation}

Finally, the total energy current flowing from site $\mu$ to site $\mu+1$ is defined as
\begin{equation}
	\label{eq:16}
	J^{(\mu,\mu+1)} = \trtxt{\hat{J}^{(\mu,\mu+1)}\hat{\rho}(t)}\,.
\end{equation}
The celebrated Fourier's law (here a discrete version) states that in a proper diffusive situation, the current inside the system is proportional to the gradient, i.e.,
\begin{equation}
	\label{eq:17}
	J^{(\mu,\mu+1)} = - \kappa [h^{(\mu+1)} - h^{(\mu)}]\,,
\end{equation}
here written in terms of energy current and energy gradient.
If both current and gradient are equal at all sites $\mu$, and furthermore, the gradient is finite, a bulk conductivity\cite{Michel2006II,Michel2006III} follows from
\begin{equation}
	\label{eq:17a}
	\kappa = \frac{J^{(\mu,\mu+1)}}{h^{(\mu)} - h^{(\mu+1)}}\,.
\end{equation}
This is called \emph{normal} or \emph{diffusive} transport.
On the other hand, if the gradient vanishes $\kappa$ diverges and the transport is called \emph{ballistic}.
However, that does not mean that the current diverges as well.
Due to the resistivity at the contact to the heat bath (in our approach $\lambda$), the current will always remain finite.

Even if we directly get a result in terms of normal or ballistic behavior for all \emph{finite} systems here, a nonzero gradient in the finite system is not sufficient to deduce normal transport in the \emph{infinite} one, too.
The influence of the contact could dominate the investigation or long ballistic waves could be suppressed in the finite system.
Thus, in order to obtain statements on the properties of the infinite system (bulk properties), it is important to investigate scaling properties as well.
For normal transport behavior, both gradient and current must tend to zero for infinitely large systems.
Then and only then the system shows diffusive behavior.
A finite current within an infinite system will always indicate ballistic transport behavior.

Note that the current operator discussed in \refeq{eq:15} is essentially the standard spin current operator, multiplied by the (large) Zeeman splitting $\Omega$.
Due to the Zeeman splitting, there is an energy flow associated with any nonvanishing spin current.
Only this energy flow is described by \refeq{eq:15}, i.e., it does not contain any energy current that would be present even if $\Omega$ was zero.
Eventually, the stationary state of the QME (\ref{eq:2}) will feature such a nonvanishing spin current.
Even though the $z$-component of the magnetization is conserved on the chain, the reservoirs may create and annihilate magnetization in $z$-direction.
Hence, we do not compensate for this current by applying an adequate magnetic gradient in the sense of Onsager\cite{Mahan1981,Zotos2005} (magnetic Seebeck effect\cite{Sakai2005}).
Doing so, for large $\Omega$, the energy flow described by \refeq{eq:15} is the dominating part of the full energy current.

%
\subsection{Monte Carlo wave-function simulation}
\label{sec:2:4}

In order to investigate the transport according to the QME requires the stationary solution $\hat{\rho}$ of Eq.~(\ref{eq:2}).
From $\hat{\rho}$ all gradients and currents can be computed with Eq.~(\ref{eq:11}) and Eq.~(\ref{eq:16}).
Unfortunately, Eq.~(\ref{eq:2}) is an $n^2$ dimensional system of linear differential equations if $n$ is the dimension of the Hilbert space.
To find the stationary state of this equation one has to diagonalize a $n^2\times n^2$ matrix which is restricted by the available memory.

A very powerful technique to find the stationary state without diagonalizing the Liouvillian is based on the stochastic unraveling\cite{Breuer2002} of the QME. The basic idea is to depart from a statistical treatment by means of density operators and turn to a description in terms of stochastic wave functions.
In fact, any QME of Lindblad form can equivalently be formulated in terms of a stochastic Schr\"odinger equation (SSE) for the wave function $\ket{\psi(t)}$
\begin{align}
	\label{eq:sse}
	\D\ket{\psi(t)}
	=& -\I\,\hat{G}(\ket{\psi(t)})\,\ket{\psi(t)}\,\D t \notag\\
	 &+ \sum_k\left(\frac{\hat{E}_k\ket{\psi(t)}}{\Vert\hat{E}_k\ket{\psi(t)}\Vert}
			-\ket{\psi(t)}\right)\D n_k\,,
\end{align}
which describes a piecewise deterministic process in Hilbert space.
The first term on the right hand side of \refeq{eq:sse} describes the deterministic evolution generated by the nonlinear operator
\begin{align}
	\label{eq:18a}
	\hat{G}(\ket{\psi(t)})
	= \hat{H}_{\text{eff}} 
	+ \frac{\I}{2}\sum_k\alpha_k\Vert\hat{E}_k\ket{\psi(t)}\Vert^2 
\end{align}
where we have introduced the non-Hermitian, effective Hamiltonian
\begin{equation}
	\label{eq:18}
	\hat{H}_{\text{eff}} = \hat{H} - \frac{\I}{2}\sum_{k=1}^4\alpha_k\hat{E}_k^{\dagger}\hat{E}_k\,.
\end{equation}
The second term in \refeq{eq:sse} contains the Poisson increments $\D n_k\in\lbrace 0, 1 \rbrace$ which obey the following statistical properties
\begin{align}
	\label{eq:18b}
	\expvalshort{\D n_k} &= \Vert\hat{E}_k\ket{\psi(t)}\Vert^2\,\D t\\
	\D n_k\,\D n_{l} &= \delta_{kl}\, \D n_k\,.
\end{align}

The stochastic process, defined by the SSE (\ref{eq:sse}), can be conveniently simulated by the following prescription. Starting from a normalized state, the first step of the unraveling procedure is to integrate the time-dependent Schr\"odinger equation according to the effective Hamiltonian defined in Eq.~(\ref{eq:18}).
Since it is not Hermitian the normalization of the state $\ket{\psi(t)}$ decreases until $\braket{\psi(t)}{\psi(t)} = \eta$, with $\eta$ being a random number drawn from a uniform distribution on the interval $\{0,1\}$ at the beginning of the step.
Subsequently, a jump $k$ takes place according to the probability
\begin{equation}
	\label{eq:20}
	p_k = \frac{\alpha_k\Vert\hat{E}_k\ket{\psi(t)}\Vert^2}
		   {\sum_k\alpha_k\Vert\hat{E}_k\ket{\psi(t)}\Vert^2}\,.
\end{equation}
Having identified the jump $k$, the state $\ket{\psi(t)}$ is replaced by the normalized state
\begin{equation}
	\label{eq:21}
	\ket{\psi(t)} 
	\rightarrow  \frac{\hat{E}_k\ket{\psi(t)}}{\Vert\hat{E}_k\ket{\psi(t)}\Vert}\,.
\end{equation}
Afterwards, the algorithm starts from the beginning again with a deterministic evolution step.
This procedure leads to one realization $r$ of the stochastic process.
Averaging over $R\rightarrow\infty$ realizations the time evolution of Eq.~(\ref{eq:2}) is reproduced\cite{Breuer2002}.

The expectation value of an observable $\hat{A}$ at time $t$ can be estimated through
\begin{equation}
	\label{eq:22}
	\trtxt{\hat{A}\,\hat{\rho}(t)}
	\approx \frac{1}{R}\sum_{r=1}^R \expval{\psi^r(t)}{\hat{A}}{\psi^r(t)}
\end{equation}
in a finite ensemble of $R$ realizations to arbitrary precision.
This is of huge practical importance, as one deals with wave functions with $\mathcal{O}(n)$ elements instead of density operators with $\mathcal{O}(n^2)$ elements.
Furthermore, if one is interested in the stationary state, ensemble averages can be replaced by time averages\cite{Molmer1996,Monasterio2007} and one single realization suffices to determine the stationary expectation value
\begin{equation}
	\label{eq:23}
	\trtxt{\hat{A}\,\hat{\rho}}
	\approx A_r = \frac{1}{T+1}\sum_{k=0}^{T} \expval{\psi^r(t_k)}{\hat{A}}{\psi^r(t_k)}
\end{equation}
with $t_k = t_0 + k\Delta t$.
It turns out that introducing this uniform time discretization, and allowing for jumps to occur at multiples of $\Delta t$ only has several technical advantages.
However, one has to bear in mind, that this introduces an error of order $\mathcal{O}(\Delta t)$\cite{Steinbach1995}.
A further problem is that for $\Delta t\rightarrow 0$, the total number of timesteps $T$ have to be increased in order to retain a sufficient number of jumps in the average (\ref{eq:23}).
This overall increase of accuracy will be purchased at the cost of tedious computations.
Nevertheless, in practice, the time-averaging procedure proves highly efficient, and results of sufficient accuracy could always be produced.
It is further advisable to discard the initial time evolution in the average in order to obtain reliable results, i.e., by choosing $t_0\gg0$.

In order to gain the standard deviation as a measure for the statistical error as well one should compute the stationary expectation value of $R$ realizations.
Thus, its mean is obtained by
\begin{equation}  
	\label{eq:24}
	\bar{A} = \frac{1}{R}\sum_{r=1}^R A_r\,,
\end{equation}
and the standard deviation of the average yields
\begin{equation} 
	\label{eq:25}
	\sigma^2 = \frac{1}{R(R-1)}\sum_{r=1}^R (A_r-\bar{A})^2\,.
\end{equation}
These errors are influenced by the chosen sampling interval $\Delta t$ the neglected steps at the beginning $t_0$ and the total amount of time steps $T$ being averaged over.
For all numerical results below we have chosen the parameters $\Delta t = 1$, $t_0=10^{4}$ and $T$ between $10^5$ and $10^6$.
For those settings the errors are surprisingly small already.

%
%
\section{Chain of Two Level Atoms}
\label{sec:3}

\begin{figure}
	\includegraphics[width=8cm]{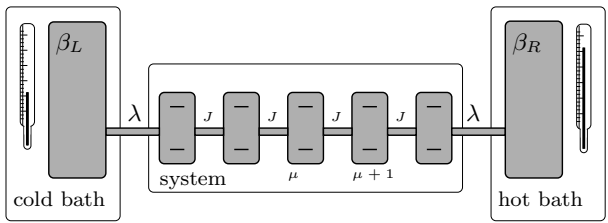}
	\caption{Chain of two-level atoms or spin-1/2 particles coupled to heat baths of different temperature.}
	\label{fig:1}
\end{figure}

First, we consider a chain of two-level atoms or spin-1/2 particles as depicted in Fig.~\ref{fig:1}.
In this case the local part of the Hamiltonian is just given by the mentioned Zeeman splitting of the individual spin
\begin{equation}
	\label{eq:26}
	\hat{h}^{(\mu)} = \frac{\Omega_{\mu}}{2}\,\Pauli{z}^{(\mu)}\,,
\end{equation}
with a splitting $\Omega_{\mu}$ which may differ from site to site.
Apart from that $\Omega$ has to be large compared to the coupling constant $J$ to remain in the weak coupling limit.
The subunits are coupled by a generalized Heisenberg interaction
\begin{align}
	\label{eq:27}
	&\hat{h}^{(\mu,\mu+1)}\notag\\
	&= \Pauli{x}^{(\mu)}\otimes\Pauli{x}^{(\mu+1)}+\Pauli{y}^{(\mu)}\otimes\Pauli{y}^{(\mu+1)}
	       + \Delta \Pauli{z}^{(\mu)}\otimes\Pauli{z}^{(\mu+1)}\,.
\end{align}
For $\Delta\neq 1$ the chain is called anisotropic chain and for $\Delta = 0$ the present model is equivalent to the XY-model (F\"orster coupling).
In this case, by plugging Eq.~(\ref{eq:26}) and the interaction given by Eq.~(\ref{eq:27}) into Eq.~(\ref{eq:15}), the current operator yields
\begin{equation}
	\label{eq:29}
	\hat{J}^{(\mu,\mu+1)}
	= \I J\Omega_{\mu}
	  \big[\Pauli{+}^{(\mu)}\Pauli{-}^{(\mu+1)}-\Pauli{-}^{(\mu)}\Pauli{+}^{(\mu+1)}\big]\,.
\end{equation}

The above given system is coupled to heat baths of different temperatures.
The left bath is set to the inverse temperature $\beta_L=0.5$, and the hotter one at the right hand side is at $\beta_R=0.25$.
Both baths couple with the same coupling strength $\lambda=0.01$ to the system.

\begin{figure}
	\includegraphics[width=8cm]{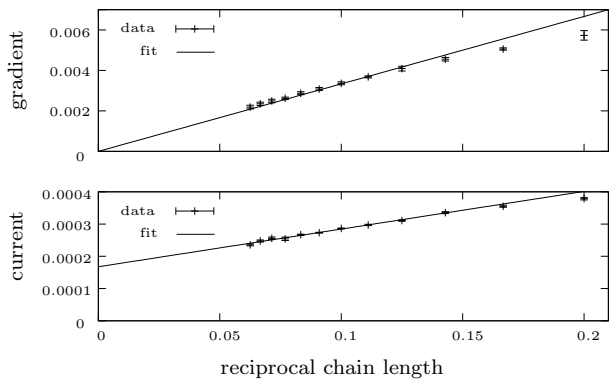}
	\caption{Scaling properties of the Heisenberg chain $N=5-16$. Lines are fits carried out for chain length $N=6-16$. System parameters: $J=0.01$, $\Delta=1$, $\Omega=1$, $\lambda=0.01$, $\beta_L=0.5$, $\beta_R=0.25$.}
	\label{fig:3}
\end{figure}

Having computed the stationary state of Eq.~(\ref{eq:2}) by using the method presented in \refsec{sec:2:4} one can compute both the stationary energy profile within the system and the current flowing through the system.
\begin{figure}
	\includegraphics[width=8cm]{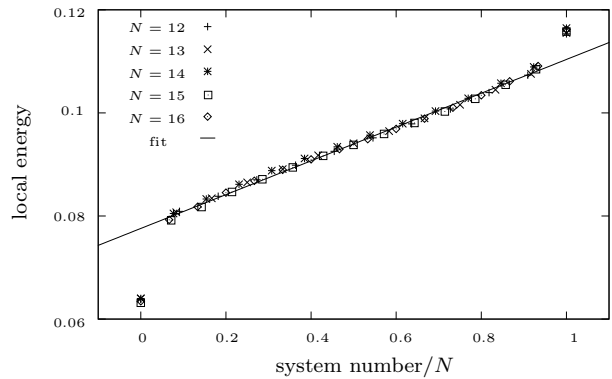}
	\caption{Local energies in a Heisenberg chain with length $N=12-16$. The system number is normalized by the chain length. The fit is carried out for chain length $N=16$ excluding site one and $16$. System parameters: $J=0.01$, $\Delta=1$, $\Omega=1$, $\lambda=0.01$, $\beta_L=0.5$, $\beta_R=0.25$.}
	\label{fig:2}
\end{figure}
In Fig.~\ref{fig:2} the internal gradient is shown for an isotropic chain $\Delta=1$ of $N=12-16$ spins according to the same constant local field $\Omega=1$ and coupling strength $J=0.01$.
To show that the gradient is equivalent for the different system sizes we have normalized the chain length in \reffig{fig:2} to one.
The fit (line in Fig.~\ref{fig:2}) is carried out for system size $N=16$ excluding the sites one and 16, because of strong influences of the contacts.
Even if the fit is done for system size 16 exclusively, all chains show the same gradient.
However, the energy difference between adjacent sites decreases for growing system sizes.
The change in the internal gradient is shown in the upper diagram of \reffig{fig:3}.
Here the gradient is plotted over the reciprocal chain length.
The error bars refer to an average over the energy differences in all adjacent pairs of sites, where the first and last pairs have been neglected again as already done in Fig.~\ref{fig:2}.
As can be seen from Fig.~\ref{fig:3} the gradient decreases for larger systems until it approaches zero for an infinite chain which is in accordance with the expected behavior in the thermodynamical limit.

The lower part of \reffig{fig:3} shows the scaling behavior of the current through the system.
Here, error bars refer to the failure produced by the stochastic algorithm given by the square root of \refeq{eq:25}.
The current decreases similar to the gradient, however, the extrapolation for the infinitely long chain does not approach zero.
A finite current for an infinite system is a typical characteristic for ballistic transport behavior.
According to the data shown in \reffig{eq:3} one could eventually conclude finding ballistic transport in the Heisenberg chain.

\begin{figure}
	\includegraphics[width=8cm]{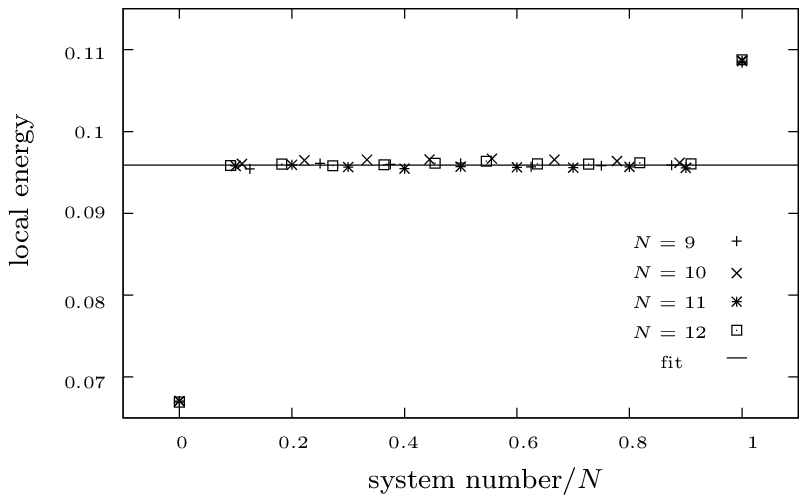}
	\caption{Local energies in a XY-chain with length $N=9-12$. The system number is normalized by the chain length. The fit is carried out for chain length $N=12$ excluding cite one and $12$. System parameters: $J=0.01$, $\Delta=0$, $\Omega=1$, $\lambda=0.01$, $\beta_L=0.5$, $\beta_R=0.25$.}
	\label{fig:4}
\end{figure}

Figure \ref{fig:4} shows the local energy profile within the XY-model ($\Delta=0$).
In comparison to Fig.~\ref{fig:2} the profile within the systems is flat.
According to  Fourier's law (\ref{eq:17}) this could be interpreted as ballistic behavior in the investigated finite models of different lengths (cf.\ discussion in \refsec{sec:2:3}).
The local current between site $\mu$ and $\mu+1$ remains finite although the gradient within the system vanishes.
This local current is constant for all investigated system sizes and we find for the chosen parameters $(5.33\pm0.05)\cdot10^{-4}$.
The results concerning the Heisenberg chain and the XY-model are in accordance with some earlier results (for smaller systems) based on the full diagonalization of the Liouvillian \cite{Saito2003,Michel2004}.

\begin{figure}
	\includegraphics[width=8cm]{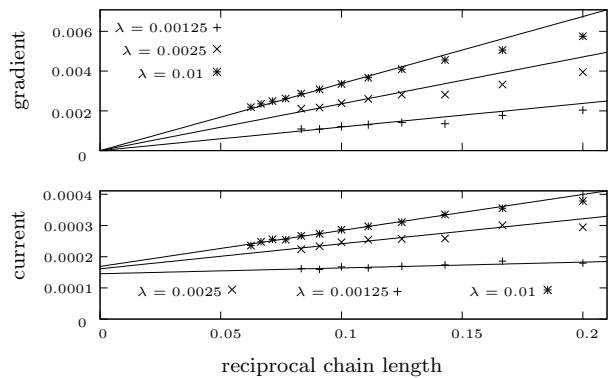}
	\caption{Dependence of the extrapolated value for the current of an infinitely long Heisenberg chain on the bath coupling strength $\lambda$.}
	\label{fig:5}
\end{figure}

In Fig.~\ref{fig:5} we investigate the dependence of the extrapolated value of the current and the energy gradient of an infinitely long chain on the coupling strength $\lambda$ at the contact.
In order to get comparable data and errors all other parameters are kept constant.
A decrease of the external coupling strength is combined with a decrease in the global decay time of the system and a drastic change of the jump probabilities (\ref{eq:20}) as well.
To gain a proper expectation value from Eq.~(\ref{eq:23}) with a rather small error it is crucial that the sampling time-step $\Delta t$ of the continuous stochastic trajectory is chosen in a way that a suitable amount of both coherent dynamics and stochastic jumps enter the average.
That means, if $\Delta t$ is too large, so that after each coherent step follows a jump already, the result of Eq.~(\ref{eq:23}) will deteriorate.
Thus, changing the external coupling strength would also require an adaption of the sampling parameter $\Delta t$.
Furthermore, in case of a larger decay time of the system also the parameter $t_0$ (initial neglect of data points) has to be increased.
Thus, having fixed these parameters to get comparable data we are restricted to a small change of the external coupling strength only.
For the finite system an increase in the external coupling strength $\lambda$ denotes that a larger current is injected into the system, as follows from Fig.~\ref{fig:5}.
The resistance of the contact is decreased.
Finally, this also results into a larger gradient within the system.
Nevertheless, Fig.~\ref{fig:5} shows that even if the results for finite systems changes drastically (especially for very small system sizes) the extrapolation for the infinite chain remains the same within the accuracy of the fit.

\begin{figure}
	\includegraphics[width=8cm]{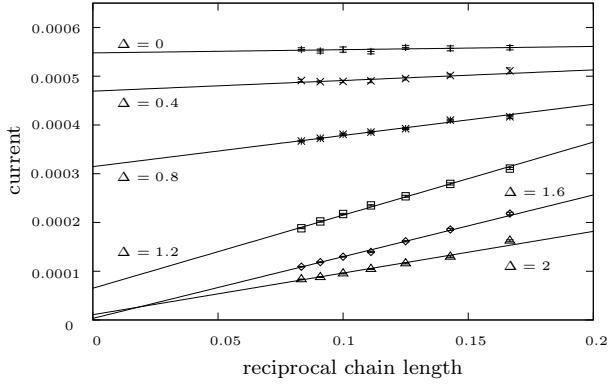}
	\caption{Scaling behavior of the current in anisotropic Heisenberg chains. System parameters: $J=0.01$, $\Omega=1$, $\lambda=0.01$, $\beta_L=0.5$, $\beta_R=0.25$.}
	\label{fig:6}
\end{figure}

\begin{figure}
	\includegraphics[width=8cm]{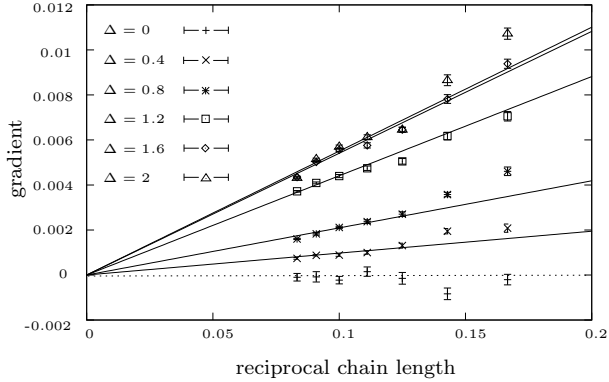}
	\caption{Scaling behavior of the gradient in anisotropic Heisenberg chains. System parameters: $J=0.01$, $\Omega=1$, $\lambda=0.01$, $\beta_L=0.5$, $\beta_R=0.25$.}
	\label{fig:7}	
\end{figure}

\begin{figure}
	\includegraphics[width=8cm]{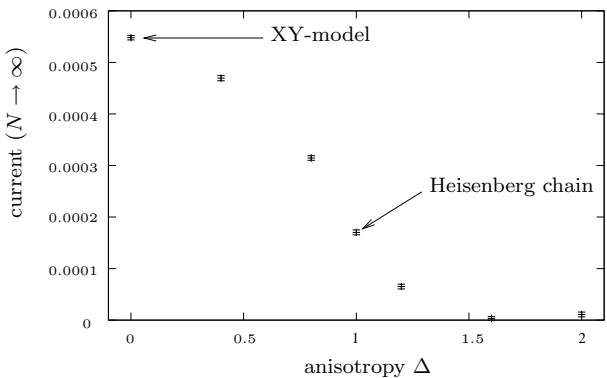}
	\caption{Extrapolated current for the infinitely long chain with dependence on the anisotropy $\Delta$.}
	\label{fig:8}
\end{figure}

Figure \ref{fig:6} shows the scaling behavior of the current for different values of the anisotropy $\Delta$ and \reffig{fig:7} shows the scaling of the gradients, respectively.
From the linear fits in \reffig{fig:6} one could extrapolate the current within an infinitely long chain.
This current is shown in Fig.~\ref{fig:8} with dependence on $\Delta$ ($\Delta=1$ refers to the Heisenberg chain and $\Delta=0$ to the XY-model).
Near the anisotropy $\Delta=1.6$ the current within the infinite system seems to vanish (cf.~\reffig{fig:6}), i.e., the analysis at hand indicates normal transport behavior.
Whether this is obtained for increasing $\Delta$ as well cannot be decided clearly from this analysis, but it seems to be probable that it remains diffusive for higher values of $\Delta$.

Having found diffusive behavior according to the Kubo formula, it seems, nevertheless, unclear how to extract the dc-conductivity from the behavior of the finite system.
Contrary to the Kubo investigation, a dc-conductivity directly follows from \refeq{eq:17a}, in the present analysis, if we assume for the moment that the linear scaling of current and gradient found in \reffig{fig:6} and \ref{fig:7} is also valid for larger systems.
According to the small errors found in the above investigation, this assumption seems to be plausible.
Thus, we are able to compute the conductivity of the infinite system for $\Delta=1.6$  using \refeq{eq:17a} by dividing the slope of the current by the slope of the energy gradient directly finding $\kappa_{\infty}=[2.34\pm0.08]\cdot10^{-2}$.
Here, the error follows from the uncertainty of the linear regression which is weighted already by the errors of the data points.

Unfortunately, the models which can be investigated according to the suggested method are also restricted in size.
The main restriction here is not the size of memory, but the time one accepts to wait for the data.
For the present technique, the computing time scales exponentially with the system size.
Thus, investigations how disorder (random offset in the local field, random couplings) would change the above results are not available yet.

\begin{figure}
	\includegraphics[width=8cm]{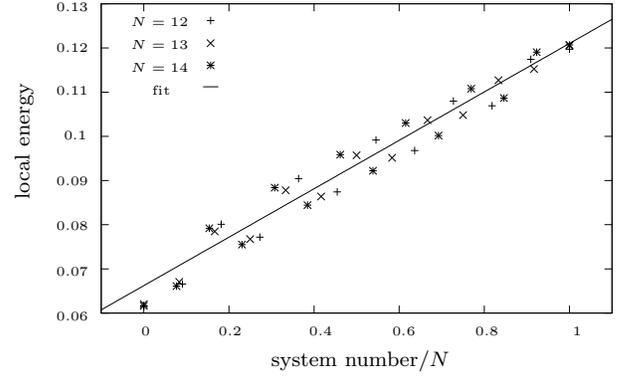}
	\caption{Local energies in an alternating local field Heisenberg chain $N=12-14$. The system number is normalized by the chain length. The fit is carried out for chain length $N=14$ excluding the first and the last site. System parameters: $J=0.01$, $\Delta=1$, $\epsilon=0.02$, $\lambda=0.01$, $\beta_L=0.5$, $\beta_R=0.25$.}
	\label{fig:9}
\end{figure}

\begin{figure}
	\includegraphics[width=8cm]{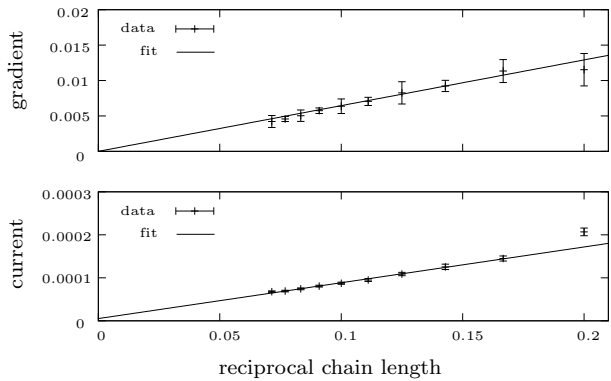}
	\caption{Scaling properties of the alternating local field Heisenberg chain $N=5-14$. Lines are fits carried out for chain length $N=6-14$. System parameters: $J=0.01$, $\Delta=1$, $\epsilon=0.02$, $\lambda=0.01$, $\beta_L=0.5$, $\beta_R=0.25$.}
	\label{fig:10}
\end{figure}

\begin{figure}
	\includegraphics[width=8cm]{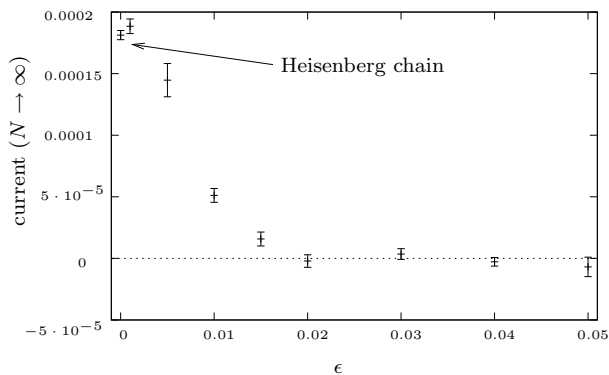}
	\caption{Current of the infinite alternating local field Heisenberg chain versus the parameter $\epsilon$ characterising the change in the local field from site to site.}
	\label{fig:11}
\end{figure}

Let us discuss another Heisenberg coupled chain of two level systems in the following.
However, we will analyze an alternating local field in the following given by $\Omega_{\mu} = 1 + (-1)^{\mu}\epsilon$ with $\mu=2,3,\ldots,N-1$, i.e., no change in the field at the edge of the system.
This keeps the contact unchanged even if $\epsilon$ is varied.
The gradient of such a system is depicted in Fig.~\ref{fig:9} and shows a modulation in the local energies according to the change in the field.
Nevertheless we have fitted the gradient by a least square fit, shown for $N=14$ in Fig.~\ref{fig:9}, and used such fits to obtain data for the gradient in systems with different size (upper part of Fig.~\ref{fig:10}).
Even if the local profile is not as flat as before the current between adjacent sites is always the same.
The scaling behavior of current and gradient is depicted in Fig.~\ref{fig:10} for $\epsilon = 0.02$.
Again the current within the infinite chain seems to vanish, which could be a hint for normal transport.
A scan over the parameter $\epsilon$ is depicted in Fig.~\ref{fig:11}.
As can be seen the current for the infinitely long chain roughly approaches zero at $\epsilon=0.02$.
For larger $\epsilon$ it remains approximately zero within the stated accuracy.
Thus, this investigation points towards normal transport behavior above $\epsilon=0.02$.

Using again \refeq{eq:17a} and dividing the slope of the current by the slope of the gradient depicted, e.g., in \reffig{fig:10} we find the conductivities of the infinite system above $\epsilon=0.02$ given in Tab.~\ref{tab:2}.
Of course this is again only correct, if the scaling behavior remains the same as already found in this finite size analysis for larger systems as well.

\begin{table}
\caption{\label{tab:2}Conductivity for the alternating local field chain in the diffusive regime.}
\begin{tabular}{ | l | l | }
	\hline
	$\epsilon$ & $\kappa_{\infty}[10^{-2}]$\\
	\hline
	0.02	   & $1.29\pm0.04$\\
	0.03	   & $0.6\pm0.3$\\
	0.04	   & $0.40\pm0.04$\\
	0.05	   & $0.09\pm0.06$\\
	\hline
\end{tabular}
\end{table}

%
%
\section{Ladder of Two Level Atoms}
\label{sec:4}

\begin{figure}
	\includegraphics[width=8cm]{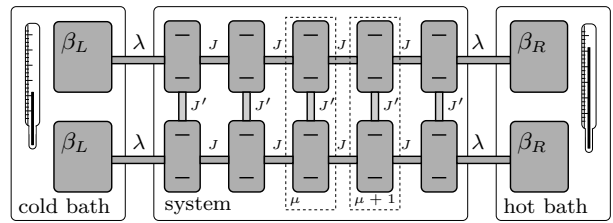}
	\caption{Ladder of two-level atoms or spins. Natural subunits are pairs of spins with an internal coupling strength $J'$. The coupling between the subunits is now given as two bonds in horizontal direction.}
	\label{fig:12}
\end{figure}

The second class of models is a spin ladder introduced in Fig.~\ref{fig:12}.
In order to consider the transport in the model, the system is partitioned into subunits $\mu$ containing two Heisenberg coupled spins with coupling strength $J'$.
Thus, the local part of the Hamiltonian for subunit $\mu$ is described by
\begin{equation}
	\label{eq:30}
	\hat{h}^{(\mu)}
	= \frac{\Omega}{2}\big(\Pauli{z}\otimes\one + \one\otimes\Pauli{z}\big)
	+ J' \vek{\hat{\sigma}}\cdot\vek{\hat{\sigma}}
\end{equation}
with the spin vector $\vek{\hat{\sigma}}=\{\Pauli{x},\Pauli{y},\Pauli{z}\}$.
Again we consider two-level systems with an energy splitting $\Omega$, here.
The interaction between two adjacent sites $\mu$ is given by
\begin{equation}
	\label{eq:31}
	\hat{h}^{(\mu,\mu+1)}
	=\sum_{i=x,y,z} \Big(\Pauli{i}\otimes\one\otimes\Pauli{i}\otimes\one+
	                   \one\otimes\Pauli{i}\otimes\one\otimes\Pauli{i}\Big)\,.
\end{equation}

Because of the weak internal couplings one may use a special type of bath contact called \emph{private bath} here.
This means that each spin at the edges of the system is coupled to its own ``private'' heat bath with temperature $\beta_L$ at the left hand side and $\beta_R$ at the right hand side.
This is different from a more general approach where the two systems at the edge are viewed as one four level system each, to which the respective heat bath couples.
The concept of private baths is only valid in the weak coupling limit again, i.e., $J,J'\ll\Omega$.

\begin{figure}
	\includegraphics[width=8cm]{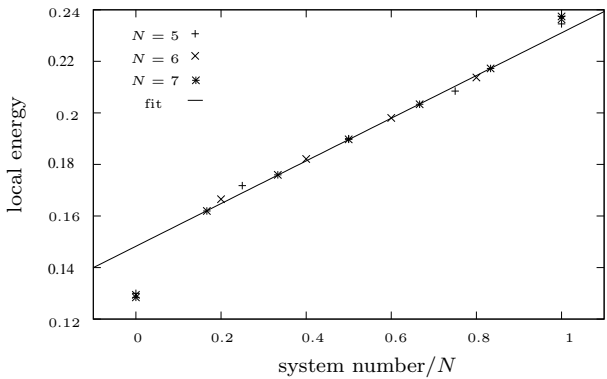}
	\caption{Local energies in a Heisenberg coupled ladder with $N=5-7$ rungs. The system number is normalized by the number of rungs. The fit is carried out for $N=7$ excluding site 1 and $7$. System parameters: $J=0.01$, $J'=0.01$, $\Omega=1$, $\lambda=0.01$, $\beta_L=0.5$, $\beta_R=0.25$.}
	\label{fig:13}
\end{figure}

\begin{figure}
	\includegraphics[width=8cm]{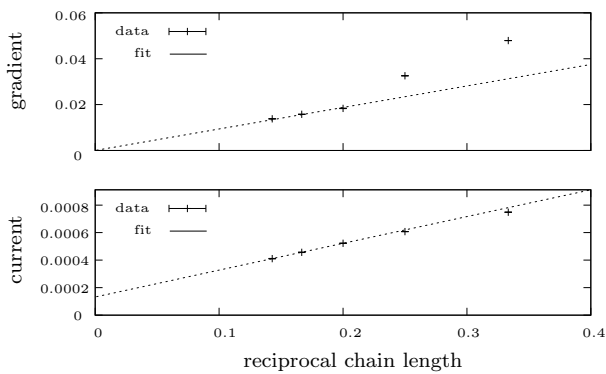}
	\caption{Scaling properties of the Heisenberg ladder with $N=2-7$ rungs. Lines are fits carried out for rungs $N=5-7$. System parameters: $J=0.01$, $J'=0.01$, $\Omega=1$, $\lambda=0.01$, $\beta_L=0.5$, $\beta_R=0.25$.}
	\label{fig:14}
\end{figure}

In Fig.~\ref{fig:13} we show the internal gradient of a ladder with $N=5-7$ rungs, according to the same coupling strength in horizontal and vertical direction $J=J'=0.01$.
Again we have normalized the length of the chain to fit all different chain lengths into the figure.
As can be seen from the figure the system features a nice linear energy gradient, again.
The scaling behavior of current and gradient for $J=J'=0.01$ is depicted in Fig.~\ref{fig:14}.
There is no possibility to compute a reasonable error for the system sizes $N=3$ and $N=4$.
This is due to the fact that for $N=3$ there is no pair without a contact to a heat bath and for $N=4$, there is just one energy difference in the center of the system away from the bath coupling.
Nevertheless, we have plotted those system sizes within Fig.~\ref{fig:14}.
The errors for larger system sizes are surprisingly small, because of the large total system size.
Due to the too small available system size and the strong influence of the baths at the edges, the gradient in the small systems is essentially different from the gradient in larger ones.
However, the current in small systems is already close to the linear fit for larger system sizes.

\begin{figure}
	\includegraphics[width=8cm]{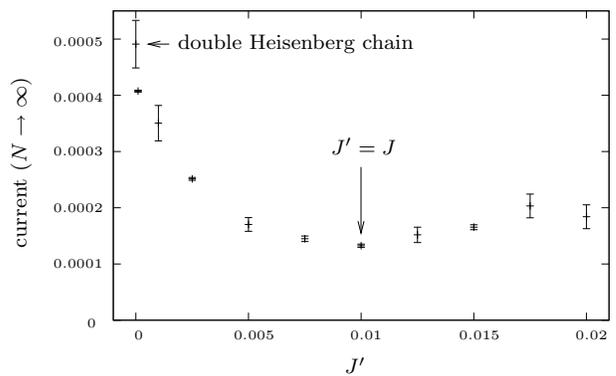}
	\caption{Current of the infinite Heisenberg ladder versus the vertical coupling strength $J'$.}
	\label{fig:15}
\end{figure}

Figure \ref{fig:15} shows the current for the infinite system extracted of the scaling analysis in dependence of the vertical coupling strength $J'$.
According to this investigation we do not find normal transport in any of the considered ladder models.
The coupling strength $J'=J$ already seems to be the minimum.
For $J'=0$ we get two completely independent Heisenberg chains with a maximum length of seven spins.
Larger values than $J'=0.02$ could be evaluated, however, results would be questionable because the weak coupling limit is violated.
Since the weak coupling limit is crucial for the derivation of the underlying QME, the results would be mathematically correct, but physically not interpretable.


\section{Conclusion}

In the present paper we have studied several models of two-level atoms or spin-1/2 particles coupled to heat baths of different temperatures.
The investigation has been based on a recently derived quantum master equation (QME) which simulates the nonequilibrium situation properly.
This QME is of Lindblad form and can, thus, be efficiently unravelled using a standard Monte Carlo wave-function technique.
The significant advantage of such an approach is its applicability to larger systems in comparison to the restricted system sizes which can be investigated by a direct diagonalization of the Liouvillian.
This follows from the fact that the stochastic unraveling deals with wave functions rather than density operators.
All interesting quantities are, here, given as mean values over stochastic trajectories and can be evaluated to arbitrary precision by adjusting the amount of timesteps averaged over.
Here, we are interested mainly in the energy profile within the system and the energy current through it.
The analysis of both current and gradient in dependence of the system size gives information on the underlying transport behavior, which could be, in principle, of ballistic or normal diffusive nature.

For finite systems, the method at hand always leads to an easily interpretable result, in terms of currents, energy profiles, and the resulting conductivities.
This is mainly a result of the concrete design of the method by a direct contact of the system with heat baths.
The extrapolation to infinite system sizes, to extract bulk properties from the analysis of finite systems, is done by a careful scaling analysis of both gradients and currents.
Finally, this provides an indication of the type of transport in an infinite probe of the model system as well.

We have analyzed a multitude of different concrete spin models here.
Those systems consist of weakly coupled two-level atoms or spins.
In the first part, we have investigated chains according to different coupling models and local fields.
The consideration was centered around the generalized Heisenberg chain, i.e., a Heisenberg chain with different ZZ-coupling strengths according to the parameter $\Delta$ within the model.
Among those, one prominent model is the XY-model with vanishing $\Delta$ featuring ballistic transport in both the finite as well as the infinite model.
Despite the relatively small system sizes investigated here, there are some hints for normal transport in the models as well:
The scaling analysis of the anisotropic model with $\Delta=1.6$ features a vanishing current for infinite chain size.
Such a finding is typically connected to diffusive behavior.
Second, the Heisenberg chain with alternating local field shows a zero current above some threshold dependent on the difference of the local fields.

Besides those investigations concerning the transport behavior, we have also considered the dependence of the results on the bath contact. 
Here, we find that the extrapolated results do not crucially depend on the respective coupling strength.

In the second part we have analysed a Heisenberg ladder with different coupling strength in horizontal and vertical directions.
According to this investigation of very short ladders, no evidence for normal diffusive behavior with dependence on the coupling strength in vertical direction is found.
The extrapolated current remains finite over the complete accessible parameter space.

In a nutshell, the comparison between our results and results based on the Kubo theory can eventually be summarized in Tab.~\ref{tab:1}.

\begin{table}
\caption{\label{tab:1}Comparison between results for the energy transport (as defined by \refeq{eq:15}) within different model systems featuring a finite local splitting $\Omega$ (finite magnetic field), as obtained from the Kubo formula and the bath coupling approach. Results for the Kubo formula are
taken from Refs.~13 and 43.} 
\begin{ruledtabular}
\begin{tabular}{  l | l   l | l }
	\textbf{Infinite Model}		& \textbf{Bath}	& $\kappa_{\infty}[10^{-2}]$ 	& \textbf{Kubo}	\\
	\hline
	\textbf{generalized Heisenberg}	& 		&				&		\\
	XY-model, $\Delta=0$		& ball.		&				& ball. 	\\
	Heisenberg, $\Delta=1$		& ball.		&				& ball. 	\\
	$\Delta<1.6$			& ball.		&				& ball. 	\\
	$\Delta=1.6$			& diff.		& {\footnotesize$(2.34\pm0.08)$}& ball. 	\\
	$\Delta>1.6$			& diff.		&				& ball. 	\\
	\hline
	\textbf{Alternating chain}	& 		&				&		\\
	$\epsilon<0.02$			& ball.		&				& ???		\\
	$\epsilon=0.02$			& diff.		& {\footnotesize$(1.29\pm0.04)$}& ???		\\
	$\epsilon>0.02$			& diff.		& 				& ???		\\
	\hline
	\textbf{Heisenberg ladder}	& ball.		&				& ???
\end{tabular}
\end{ruledtabular}
\end{table}

The heat bath coupling approach to transport behavior presented in this paper features some significant advantages:
Besides the direct determination of current and gradient in the finite system, and thus, of the conductivity in any concrete finite situation, the method also allows us to extract the conductivity of the infinite model by means of an extrapolation.
This extrapolation relies on the linear scaling behavior of current and gradient toward zero, which is extracted from the analysis of finite system sizes.
Thus, the direct coupling of reservoirs to model systems substantially improves the understanding of the transport behavior of such models on both small and infinite scales.

\begin{acknowledgments}
We thank H.-P.\ Breuer and M.\ Henrich for fruitful discussions concerning the QME, and J.~Hamm and K.~B\"ohringer for their help with the computing system. Financial support by the Deutsche Forschungsgemeinschaft through the ``Graduiertenkolleg 695'' is gratefully acknowledged.
\end{acknowledgments}

\end{document}